\documentclass{article}
\usepackage{graphicx} 
\usepackage{hyperref}
\hypersetup{
    colorlinks=true,
    allcolors=black
}
\usepackage{braket}
\usepackage{titlesec}
\usepackage{geometry}
\usepackage{minted}
\usepackage{amsmath}
\usepackage{xcolor}
\usepackage{comment}
\usepackage{quantikz}

\newcommand{\ec}{\gate[style={draw,minimum width=0.55cm,minimum height=0.45cm}]{\text{\scriptsize EC}}}
\newcommand{\leftlabel}[2]{\llap{\smash{#1}\quad}#2}

\usepackage{subfig}

\usepackage[backend=bibtex, style=numeric, sorting=none]{biblatex}
\title{Developments in superconducting erasure qubits for hardware-efficient quantum error correction}
\author{Maria Violaris\thanks{ mviolaris@oqc.tech}, Luciana Henaut, James Wills, Gioele Consani, Jamie Friel, Brian Vlastakis\\Oxford Quantum Circuits}

\date{January 6, 2026}

\begin{document}

\maketitle

\begin{abstract}
Quantum computers are inherently noisy, and a crucial challenge for achieving large-scale, fault-tolerant quantum computing is to implement quantum error correction. A promising direction that has made rapid recent progress is to design hardware that has a specific noise profile, leading to a significantly higher threshold for noise with certain quantum error correcting codes. This Perspective focuses on erasure qubits, which enable hardware-efficient quantum error correction, by concatenating an inner code built-in to the hardware with an outer code. We focus on implementations of dual-rail encoded erasure qubits using superconducting qubits, giving an overview of recent developments in theory and simulation, and hardware demonstrators. We also discuss the differences between implementations; near-term applications using quantum error detection; and the open problems for developing this approach towards early fault-tolerant quantum computers.
\end{abstract}

\section{Introduction}

Implementing quantum error correction (QEC) for fault-tolerant quantum computers is a key challenge for today's quantum industry. To be commercially useful, quantum computers require the capability to run algorithms at high speed, with high-quality qubits, at scale. The global quantum community is now working to achieve quantum computers capable of millions, billions and trillions of quantum operations, with each of these regimes unlocking new classes of useful applications \cite{preskill2025beyond,OQC_Technical_Roadmap, UK_DSI_NQSM_2023,IBM_Quantum_Technology_Roadmap,Quantinuum_Roadmap_2024,Gidney_2025_RSA_Under_Million,riverlane2024quantum}. Achieving the error-rates required for these regimes with standard QEC approaches requires huge numbers of physical qubits \cite{Fowler2012SurfaceCodes, Riverlane2025QEC}.\\

There has been significant progress in addressing the challenge of building high-quality logical qubits using QEC. This includes advances in the theory of error-correcting codes, and milestone experimental demonstrators that have validated the technical feasibility of primitives required for QEC \cite{roffe2019quantum, bluvstein2024logical, da2024demonstration, zhao2022realization, google2024quantum}. However, the number of physical qubits required for achieving logical qubits with useful error-rates remains a prohibitively large engineering challenge. It is therefore essential to drastically reduce qubit counts and hardware overheads to access the useful applications of quantum computers in realistic timescales. One of today's key challenges when building hardware for quantum error-correction is therefore to ensure that hardware overheads do not scale exponentially with increased system size, and instead remain at manageable scales.\\

A promising avenue to solve the scaling challenge is to design hardware specifically for quantum error correction, enabling the low error-rates required for useful quantum computation to be reached with hardware parameters that are realistic and achievable in the near future.  Designing hardware that reduces error-correction overhead is known as \textit{hardware-efficient error-correction}. The principle behind this approach is to design hardware that takes advantage of the native physics of the device within the error-correction protocol. Then target logical error-rates can be reached with minimal hardware footprint and complexity. \\

Hardware-efficiency can encompass a broad range of hardware overhead savings, such as reduced physical qubit count; reduced hardware footprint; increased error-correction noise thresholds; and reduced requirements on control and cryogenic infrastructure.  In recent years, hardware design explicitly tailored for error correction has become an increasingly popular and widespread strategy, giving rise to a range of hardware-efficient QEC approaches that can access aspects of these hardware savings \cite{wu2022erasure, kubica2023erasure, teoh2023dual, scholl2023erasure, kang2023quantum, koottandavida2024erasure,gu2023fault, levine2024demonstrating, hann2024hybrid, guillaud2021error, guillaud2019repetition, xu2023autonomous}. A common feature is to engineer qubits such that their dominant noise-source can be corrected more efficiently. This works by concatenating the qubits with a specially selected code that corrects the dominant physical noise, resulting in significantly better error thresholds for fault-tolerance, and reducing the number of physical qubits required for achieving a target logical error rate.\\

For a truly hardware-efficient system, we must engineer advantageous noise profiles on qubits without adding prohibitive complexity in the hardware infrastructure required to implement them. The approach to hardware-efficient QEC that we focus on here is erasure qubits, which result in errors that has known locations. A central benefit of this noise structure is that the system can reliably flag when and where an error has occurred.\\

Conventional QEC is designed to correct errors whose locations are unknown, which is inherently harder and demands large hardware overheads. In contrast, erasure qubits provide explicit location information that can be incorporated into the decoder — the classical algorithm that interprets syndrome measurements. Knowing the error locations in advance allows quantum codes to tolerate much higher physical error rates than in the standard Pauli-noise setting. This leads to orders of magnitude reductions in logical error rates, together with substantial increases to threshold for many codes \cite{gu2024optimizing, chang2024surface, gu2023fault}. These novel qubits can be engineered using superconducting circuits, and in principle require minimal or no increase to hardware infrastructure complexity \cite{wills2025error}. \\

In this Perspective, we outline the key concepts underlying erasure-based quantum error correction and explain why erasure qubits provide a powerful route to hardware efficiency. We introduce the relevant QEC and superconducting-circuit background before examining how erasure noise arises, how dual-rail encodings convert amplitude damping into detectable erasures, and how this leads to improved logical error rates and higher noise thresholds. We then review recent theoretical and experimental progress, across both academic and industrial groups, in analysing, simulating, and implementing superconducting erasure qubits. Beyond long-term fault tolerance, we also highlight how erasure-aware primitives can serve as practical tools for error detection and mitigation on near-term devices. Finally, we discuss the open challenges and opportunities in engineering erasure-biased noise and integrating it with advanced decoding and code-design strategies.

\subsection{QEC background}

Quantum error correction works by using a number of physical qubits to encode a logical qubit, which is then protected against some number of errors on the physical qubits. A quantum error correction protocol enables one to retrieve information about what type of errors occurred in the physical qubits, and in which locations, such that the physical errors can be corrected. This process ensures the computational information encoded in the logical qubits remains protected \cite{NielsenChuang, Gottesman1997, PreskillNotes}. \\

In classical error correction, the information about what errors occurred to physical bits encoding a logical bit can be retrieved by measuring the bits and comparing them. By contrast, quantum error correction has the added complication that standard measurements in the computational basis will collapse the physical qubits, causing unwanted decoherence that destroys the computation. Hence, physical qubits cannot be directly measured to identify what errors occurred. \\

\begin{figure}
    \centering
    \includegraphics[width=1\linewidth]{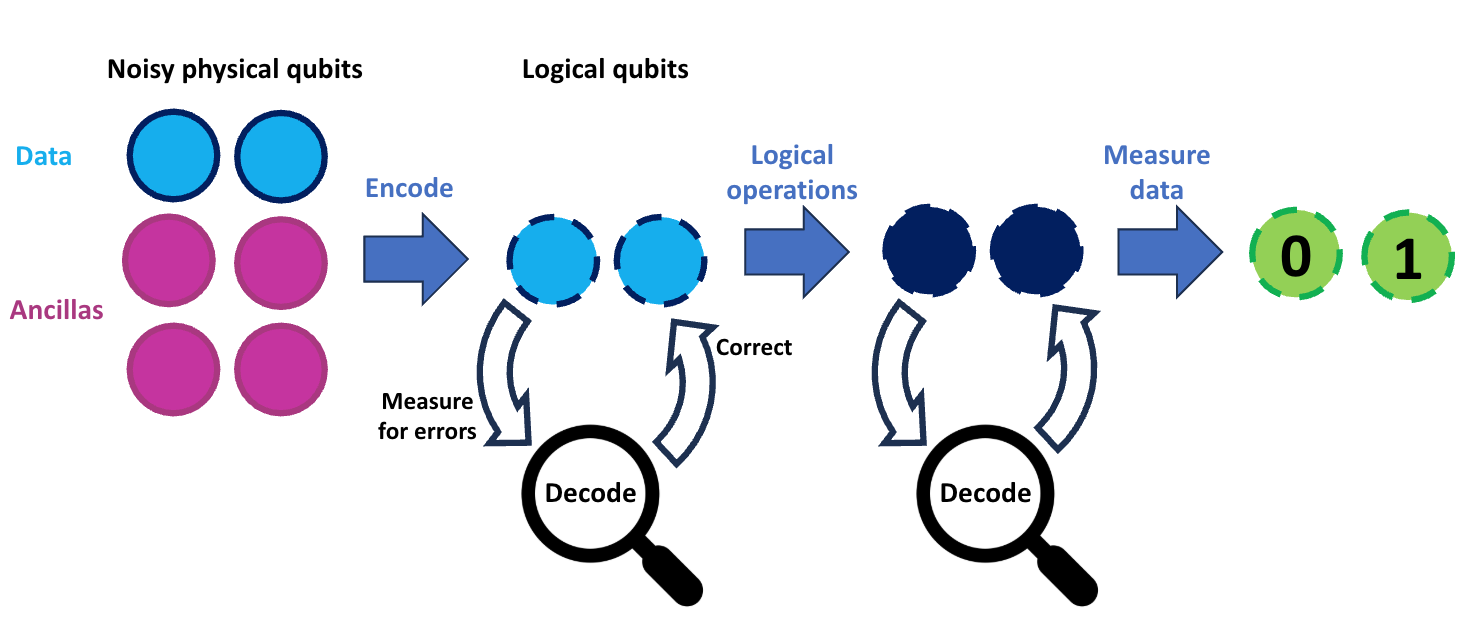}
    \caption{A visualisation of the quantum error correction process, which involves encoding data from physical qubits; frequent measurements for errors, decoding, and correction, while applying logical operators to the encoded information; and finally logical measurement to reveal the computational information.}
    \label{fig:qec}
\end{figure}

Despite this limitation, it was shown by Calderbank and Shor \cite{calderbank1996good}, Steane \cite{steane1996multiple}, and in subsequent studies that quantum error correction is in fact possible. This is by designing measurements that extract information about what errors happened and where (the \textit{syndrome}), but also project the noisy physical qubits into a correctable subspace. The resulting state of the physical qubits is then related to the correct state by a series of known gates that are revealed by the syndrome. The appropriate pattern of gates is deduced by an algorithm which estimates the most likely pattern of errors, the process of \textit{decoding}. After applying the gates identified by the decoder, the correct, coherent logical information is restored. The overall process of applying quantum error correction is visualised in figure \ref{fig:qec}.\\

Error-correction alone is insufficient for a code to enable fault-tolerant quantum computing; it also requires a threshold theorem, i.e. a combination of error-correcting code, physical noise model and decoding algorithm such that increasing the code distance reduces the logical error rate. A theoretical breakthrough came when this was proven to be theoretically possible by multiple groups \cite{aharonov1997fault, knill1998resilient, kitaev2003fault}. Demonstrating quantum error correction below threshold has become a key aspect of recent experimental milestones demonstrating that quantum error correction works in practice, for instance in Google Quantum AI's demonstration of below-threshold error correction \cite{google2024quantum}.\\

Two standard noise models are commonly considered in this context: the \textit{circuit noise} model and the \textit{phenomenological noise} model \cite{dennis2002topological, fowler2009high, wang2011surface}. In the circuit noise model, errors can occur at any stage of the QEC cycle. By contrast, the phenomenological noise model introduces errors directly on data qubits and measurement outcomes at fixed rates, offering a greatly simplified abstraction. Despite its simplicity, the phenomenological model remains valuable for theoretical studies, as it captures the essential challenges of fault-tolerant quantum computing in a tractable form. To approximate more realistic noise processes within this framework, Pauli twirling techniques~\cite{twirling} are often employed, enabling the reduction of general noise channels to effective Pauli channels compatible with stabilizer simulation. 

\subsection{Superconducting circuits background}\label{subsec:superconducting-background}

Superconducting qubits are a widely-adopted quantum computing modality due to their advantages in manufacturing, fast gate-speeds and flexibility in design. The materials and processes involved in manufacturing are adapted from well-established techniques in the semiconductor industry, while integration with microwave control and readout hardware leverages decades of progress in superconducting electronics, telecommunications and radio-frequency (RF) engineering \cite{krantz2019quantum}. \\

The core components of superconducting qubits are called \textit{Josephson junctions} (JJs), which exhibit a macroscopic quantum effect arising from quantum tunnelling \cite{martinis2004superconducting, tafuri2019fundamentals}. Often referred to as an `artificial atom', the emergent behaviour of the JJs parallels the structure of atomic energy levels \cite{martinis1985energy}. The excitation of these levels can be controlled using microwave pulses to create superconducting qubits.  \\ 

By innovating on superconducting circuit designs, these artificial atoms can be engineered to have specific properties. For example, JJs can be combined with other circuit elements such as capacitors or inductors. The most prevalent design is the transmon qubit, which has become widespread in academic research and in industry \cite{Koch2007Transmon}. The transmon circuit includes a large shunt capacitor, which reduces charge noise sensitivity while retaining the anharmonicity needed to operate gates. Other superconducting qubits include the flux qubit, fluxonium, and phase qubits, each with different noise trade-offs \cite{Mooij1999FluxQubit,Manucharyan2009Fluxonium,Martinis2002PhaseQubit}. \\

\textbf{Noise sources in superconducting qubits}\\

Superconducting qubits experience several forms of noise arising from their interaction with the environment and from imperfections in their underlying materials. Noise within the computational subspace $\{\ket{0}, \ket{1}\}$ broadly takes two forms. One is \emph{energy relaxation} or \emph{amplitude damping}, where the qubit decays from the excited state $\ket{1}$ to the ground state $\ket{0}$, which can be modelled as a Pauli-$X$ type error. The second is \emph{dephasing}, where the relative phase between $\ket{0}$ and $\ket{1}$ becomes randomized, corresponding to a Pauli-$Z$ type error. The characteristic timescales for these processes are denoted $T_1$ and $T_2$. Materials defects, interface losses, and junction variability are known to limit both parameters in superconducting devices \cite{Oliver2013Coherence}.\\

In addition to errors that occur within the computational subspace, superconducting qubits can also suffer from \emph{leakage noise}, where the qubit state leaves the two-level computational basis and occupies a higher excited state such as $\ket{2}$. Because leaked states fall outside the error model of most fault-tolerant schemes, they can persist across multiple error-correction cycles and propagate correlated errors through entangling gates \cite{Peterer2015HigherLevels,Martinis2002PhaseQubit}. Leakage can be induced by control imperfections, for example when strong microwave drives inadvertently populate higher transmon levels. Mitigation strategies include active leakage-reduction units that return leaked population to the computational subspace \cite{PRXQuantum.2.030314}, and qubit designs with larger anharmonicity (such as fluxonium) that inherently suppress leakage \cite{Manucharyan2009Fluxonium}.\\

The underlying physical mechanisms responsible for relaxation, dephasing, and leakage are closely tied to materials science and fabrication. Surface losses, interface contamination, dielectric defects, and junction-barrier variability can strongly influence coherence times and overall device reliability. Recent experimental work in industry has begun to probe these mechanisms directly, including studies of coherence metrology in multimode devices, analysis of Josephson-junction barrier variation, and substrate-engineering methods aimed at reducing material-induced noise \cite{wills2025error,OQC2025JJBarrier,OQC2025Substrate}.

\subsection{Trends in superconducting QEC}\label{subsec:trends}

The dominant quantum error-correcting code for superconducting architectures is the surface code, favoured for its relatively high threshold and compatibility with strictly local, nearest-neighbour connectivity \cite{dennis2002topological}. The compatibility of the code with local connectivity is especially important for superconducting qubits, which natively interact only with adjacent neighbours set by the lattice geometry. The surface code’s threshold is often quoted at around 1$\%$, a physical error rate that makes fault-tolerant quantum computing with this code plausibly attainable \cite{barends2014superconducting}. Due to these properties, the surface code is the most well-researched and widely adopted approach to QEC. It is also the code that Google Quantum AI used for their demonstrations of the full quantum error correction cycle below threshold \cite{google2024quantum}. However, it has the drawback of requiring a large number of physical qubits per logical qubit. \\

In recent years, there has been increasing interest in \textit{high-rate qLDPC codes} which can have a higher encoding rate of logical qubits per physical qubit than the surface code \cite{breuckmann2021quantum}. For example, IBM Quantum published research into the bivariate bicycle code \cite{bravyi2024high, yoder2025tour}. These codes have potentially large savings in qubit count, though for superconductors there is a hardware challenge of implementing long-range gates. While these codes typically require significant hardware modifications for superconductors, a recent proposal from Riverlane researchers identifies a new family of high-rate qLDPC codes known as \textit{directional codes}, which can be implemented on local connectivity \cite{geher2025directional}. Unlike the surface code, there are not standard methods for implementing fault-tolerant logical operators with qLDPC codes, though IBM researchers recently published progress in this direction \cite{yoder2025tour}. The 2024 and 2025 Riverlane QEC reports therefore summarised qLDPC codes as a ``high risk, high reward" approach to quantum error correction \cite{riverlane2024quantum, Riverlane2025QEC}. \\

There are many other codes with their own benefits and drawbacks that are attracting significant research interest, including color codes \cite{Bombin2006ColorCodes,Landahl2011ColorCodes} and Floquet codes \cite{Hastings2021Floquet,Fahimniya2025HyperbolicFloquet}.
\\

\textbf{Co-designing hardware noise with QEC codes}\\

Recent advances in both quantum hardware and quantum error-correction theory have deepened the interplay between the two, making co-design a central theme of current research and progress \cite{eisert2025mind}. The standard noise model used to investigate quantum error correcting codes is Pauli noise, which can be decomposed into Pauli X, Y and Z-errors in the computational subspace. This is a simplification of the actual noise on hardware. Approximating noise as Pauli noise is useful for benchmarking quantum error correcting codes, while accurately modelling hardware errors requires more involved analysis and simulation techniques. As we will discuss, co-design of error-correcting codes and native noise profiles is the design principle behind superconducting erasure qubits, enabling both a higher encoding rate and increase noise threshold. \\

Other efforts to engineer noise of superconducting qubits include bosonic encodings such as cat qubits, which realize strongly biased Pauli noise. At suitable operating points, bit-flip processes are exponentially suppressed relative to phase-flip processes \cite{mirrahimi2014dynamically, tuckett2018ultrahigh, roffe2023bias}. This biased Pauli noise can be leveraged by tailored decoders and codes, so the code concentrates on the dominant error channel, increasing threshold and encoding rate. The approach is being actively explored by academic groups (e.g. Yale, Caltech) and industry (e.g. Alice $\&$ Bob, AWS), with small-scale demonstrations exhibiting the requisite biased noise \cite{reglade2024quantum, hann2024hybrid, putterman2025hardware}. Theoretical thresholds and physical qubit resource savings under Pauli biased noise have been analyzed for several code families \cite{mirrahimi2014dynamically, tuckett2018ultrahigh, roffe2023bias}. Another bosonic approach is the GKP (Gottesman-Kitaev-Preskill) code, which converts continuous displacement noise into X- and Z-type error syndromes. Employed as an inner code, GKP can substantially lower qubit-number overhead via concatenation with outer codes such as the surface or XZZX code \cite{Conrad2022GKPcodes, Zhang2023ConcatenationGKPXZZX}. \\

The flexibility of hardware errors to deviate from standard Pauli noise models is therefore not only a challenge, but an opportunity: quantum error-correcting codes can be adapted to take advantage of the unique features of hardware noise, and qubits can be intentionally designed to exhibit errors that can be corrected more efficiently. 

\section{Erasure noise}

The dominant errors on erasure qubits cause a qubit's state to leave the standard computational subspace, followed by a measurement that detects this error. When an error that leaves the computational subspace is detected, it is termed a \textit{heralded erasure} or \textit{erasure error}. By contrast, an error that leaves the computational subspace and is not detected is leakage, a harmful source of noise introduced in section \ref{subsec:superconducting-background}. Erasure errors are leakage errors that are engineered to be detectable, enabling powerful error-correction properties. In this section we explain how erasure qubits can be implemented using dual-rail encoding, and review analytic and numerical results on how they increase the power of quantum error-correcting codes. 

\subsection{Dual-rail encoding}\label{subsection:erasure-conversion}

The most common way to implement erasure qubits on superconductors, and other hardware modalities, is using \textit{dual-rail encoding}. This form of encoding originated in photonics, and was important for demonstrations of the feasibility of a fault-tolerant photonic quantum computer \cite{knill2001scheme}. \\

A dual-rail qubit encodes a single logical qubit in the Hilbert space of two physical qubits. We can consider the 0-logical state $\ket{0}_L = \ket{10}$, and 1-logical state $\ket{1}_L = \ket{01}$. Then, a physical de-excitation from any state $\ket{\Psi}_L = \alpha \ket{0}_L + \beta \ket{1}_L = \alpha \ket{10} + \beta \ket{01}$ to $\ket{00}$ results in a qubit state that is outside of the logical computational subspace, $\{\ket{0}_L, \ket{1}_L\}$. This means we can in principle detect the erasure error without causing a collapse of the logical computational subspace, by determining only whether or not the system is in a fully de-excited state, rather than which computational state it is in. This is possible by performing measurements (\textit{erasure checks}) that only distinguish the $\{\ket{10}, \ket{01}\}$ and $\{\ket{00}, \ket{11}\}$ physical qubit subspaces. Note that in practice, implementations typically have $\ket{1}$ being the excited state and $\ket{0}$ being the ground state, such that de-excitations to $\ket{00}$ are the dominant error, while excitations to $\ket{11}$ are suppressed. Erasure checks in practical implementations will be imperfect, with some probability of false-positive and false-negative results, that must be accounted for when assessing the practicality of schemes with dominant erasure noise \cite{teoh2023dual, chang2024surface}.\\

\begin{figure}
    \centering
    \includegraphics[width=0.5\linewidth]{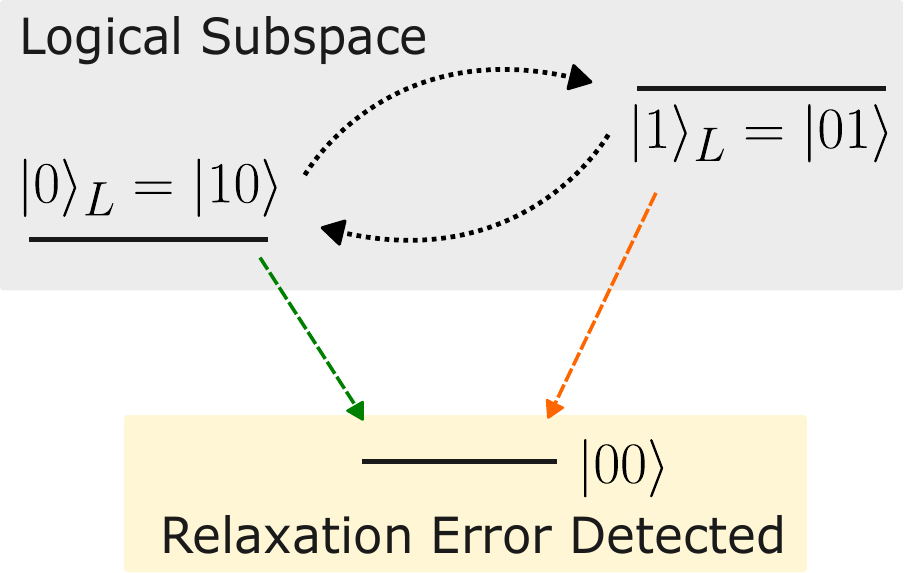}
    \caption{A visualisation of dual-rail encoding, with arrows indicating the suppressed transitions between states in the logical subspace, and dominant transitions to the ground state (a detectable erasure error).}
    \label{fig:dual-rail}
\end{figure}

When dual-rail encoding was applied to superconducting circuits, the term \textit{erasure qubits} was coined for qubits in which erasures are (purposely designed to be) the dominant error source — a form of biased noise \cite{kubica2023erasure}. The de-excitation of a dual-rail qubit from one of the computational logic states to the ground state is an erasure error, as depicted in figure \ref{fig:dual-rail}. Using erasure qubits and dual-rail encoding is now an effective and widespread approach towards implementing error mitigation and correction, extending beyond photonics to neutral atoms \cite{ma2023high, chow2024circuit, wu2022erasure, baranes2025leveraging, pecorari2025quantum}, trapped ions \cite{quinn2024high, shi2025long, kang2025doubling, kang2023quantum}, and as detailed in section \ref{section:implementations}, superconducting qubits.

\subsection{Hierarchy of errors} \label{subsec:hierarchy}

We now consider erasure errors within the hierarchy of how harmful an error is. The most harmful type of error is leakage, as it is undetectable and propagates into correlated errors that are not flagged by the decoder. Next is standard Pauli noise. Arbitrary noise channels within the computational subspace can be modelled as Pauli noise channels for the purpose of quantum error correction, and this kind of noise can be corrected by discerning the likely error type and location from the error-correcting code.  Finally the least harmful of these three error types is an erasure error, because the location of the error is heralded. The decoder can then identify and correct the error much more efficiently than without location information. \\

Given this error-hierarchy of how easy errors are to correct, we can deduce the ideal noise profile that we would like to engineer in our qubits for optimal error correction, with the dominant noise source being the easiest to correct. This requires erasure errors to be dominant, with a suppressed number of Pauli errors, and an even more suppressed presence of leakage. This is illustrated with an $n=4$ surface code example in figure \ref{fig:error-types}. Engineering this error-hierarchy is therefore important in order to maximise the effectiveness of an error-correcting code. 

\subsection{Scaling of erasure error-correcting codes}

Directly retrieving the location information of dominant errors is a tantalizing prospect, with powerful results showing potential for increased physical error thresholds and sub-threshold scaling of logical error rate. These advantages originate from the improvement in code distance given by location information. \\

Code distance is defined as the minimum weight error that will not be detected by the code, with \textit{weight} being the number of qubits that are affected by the error. In \cite{grassl1997codes}, it was shown that a code of distance $d$ can correct any $d-1$ or fewer erasure errors, compared with the standard limit of $\left\lfloor \frac{d-1}{2} \right\rfloor$ arbitrary errors. The former follows directly from the fact that $d-1$ errors can be detected in a distance $d$ code, with correction being possible with supplementary location information. Hence there is a significant gain in the number of errors that can be corrected for a given code when including information about the location of errors. An illustrative example is that while the smallest size code for correcting arbitrary errors requires five qubits, adding the location information of erasure errors reduces the minimum qubit number to four \cite{grassl1997codes} (see section \ref{subsec:surface-example} for an example of this result with the surface code). This demonstrates how an erasure noise model enables a given number of physical qubits to have enhanced error-correction capabilities.\\

Subsequent work showed that incorporating erasure-location information can dramatically raise the fault-tolerance threshold of a code \cite{stace2009thresholds, barrett2010fault, knill312190scalable}. In the idealised limit of perfect syndrome measurements (the \textit{code-capacity} noise model), the surface-code threshold with location information reaches $50\%$, whereas without location information the corresponding depolarising-noise threshold is upper-bounded at around $19\%$ \cite{Wootton2012HighThreshold, Bombin2012StrongResilience}. This $50\%$ value saturates the fundamental bound imposed by the no-cloning theorem.\\

In analyses of surface codes that convert amplitude-damping events into heralded erasures while leaving the remaining noise Pauli-like \cite{kubica2023erasure}, simulations show a 5.2x factor improvement in the Pauli noise error rate that can be tolerated when using a code based on erasures, compared with a protocol where these errors go undetected. More generally, logical error probability scales more favourably with code distance under erasure noise than under Pauli noise. \\

To see this, we can compare the expected scaling of logical error probability $p_L$ with erasure noise $e$ and Pauli noise $p$, following the heuristic in \cite{gu2024optimizing}. Let $e^{*}$ and $p^{*}$ be the threshold probabilities under erasure noise and Pauli noise respectively, with $d$ being the code distance. For erasure noise,
\begin{align*}
p_L &\propto \left(\frac{e}{e^{*}}\right)^{d},
\end{align*}
because an uncorrectable error requires $d$ heralded erasures to occur, whereas under Pauli noise
\begin{align*}
p_L &\propto \left(\frac{p}{p^{*}}\right)^{\left\lceil d/2 \right\rceil},
\end{align*}
since ${\left\lceil d/2 \right\rceil}$ undetected faults suffice to mislead the decoder \cite{gu2024optimizing}.\\

\subsection{Advances in erasure QEC theory and simulation }

Erasure noise has been demonstrated in circuit-level noise simulations and theoretical analysis to enable increased threshold and reduced logical error probability in the surface code, with one study analysing frequency of erasure checks finding an approximately 50$\%$ increase in threshold  \cite{gu2024optimizing} and another focused on imperfect erasure checks finding threshold at least over twice that of Pauli noise \cite{chang2024surface}. Gains in threshold and sub-threshold scaling have also been found for Floquet codes \cite{gu2023fault}. \\

In the case of high-rate quantum LDPC codes, a study of the La-cross code and bivariate bicycle code found orders of magnitude reductions in logical error probability \cite{pecorari2025quantum}. The threshold improvement of the La-cross code with erasures was similar to the surface code, though the bivariate bicycle code did not show significant threshold improvements. Numerical results therefore suggest that the magnitude of increase in threshold depends on the details of the code, while significant improvements to sub-threshold scaling and logical error probability apply for arbitrary codes.  In general, it is expected that erasure qubits will benefit any code they are concatenated with, though finding which codes benefit most from erasure qubits and characterising impacts on threshold and logical errors is an active area of research \cite{gu2023fault}.\\

A further advantage comes from \textit{biased} erasure noise, where not only the location of the error but also the specific state involved in the decay is known. This scenario arises naturally in architectures where only one of the dual-rail basis states undergoes amplitude damping. The resulting erased qubit can then be reset back to the computational state from which it decayed, creating a bias in the effective Pauli errors in the qubits, which can be exploited by the decoder \cite{sahay2023high}. Recent theoretical studies \cite{sahay2023high, wu2022erasure} show the significant improved performance that such biased erasures can achieve relative to symmetric erasures. In particular, thresholds can increase by factors of approximately 2–5, depending on the bias ratio and code family, and logical error rates at fixed physical error can be correspondingly reduced.\\

A recent study additionally found significant benefits to using erasure qubits for implementing non-Clifford gates, with both analytical results and numerical simulations \cite{jacoby2025magic}. They considered magic-state distillation using the surface code. A distillation protocol requires first using magic-state injection circuits to prepare low-error magic states, that are then used for distillation to lower-error states. The injection process involves error-detection, and postselection on the states without errors. In simulations of magic-state injection protocols with erasure qubits, they were found to have significantly lower logical error rates than with the standard Pauli noise. Meanwhile, the acceptance rate was only marginally reduced. Similarly, logical error rate savings were also found for the cultivation step of the magic-state cultivation protocol, which was recently proposed to prepare high-fidelity magic states at substantially lower overhead than traditional distillation \cite{gidney2024magic}.\\

Based on this, the authors of \cite{jacoby2025magic} estimated that erasure qubits used together with magic state cultivation can achieve the error-rate required for early fault-tolerant quantum applications with physical (heralded) erasure error rates less than around $4 \times 10^{-3}$, and residual Pauli error rates of order $10^{-4}$. This demonstrates an important tipping point for superconducting qubits: the physical error-rates required for low logical error rates are becoming realistically achievable using near-term hardware.\\

\subsection{Advantage of erasures for superconducting qubits}\label{subsec:erasure-advantage}

The application of erasure qubits is particularly notable for superconductors, given that they promise to bypass the typical limiting factor of physical superconducting qubit fidelity, namely the T$_1$ error from amplitude damping of the excited physical state $\ket{1}$ to the ground state $\ket{0}$\cite{kubica2023erasure}. These physical qubit fidelities are the limiting factor for using superconducting qubits in QEC, especially since the other dominant error source, which comes from dephasing, can be significantly reduced with known techniques such as dynamical decoupling. There have been significant recent developments in proposals and demonstrations of erasure qubits with superconductors, including work at OQC \cite{wills2025error} and multiple other academic, industry and collaborative groups \cite{kubica2023erasure, teoh2023dual, chou2024superconducting, koottandavida2024erasure, levine2024demonstrating}. \\

The principle of co-designing the noise profile of qubits with an error-correcting code applies for both erasure qubits and other strategies such as biased Pauli noise, as is the case for instance in cat qubits (discussed in section \ref{subsec:trends}). Both enable an increased ratio of logical-to-physical qubits, aiding scalability, however they have distinct practical trade-offs. While biased Pauli noise enables in principle very high logical-to-physical qubit ratios, engineering the biased noise on these qubits requires additional control wiring, creating an infrastructure scaling challenge. By contrast, engineering bias towards erasure noise does not necessitate increased complexity in control infrastructure. This is illustrated by the multimode qubit implementation. In a coaxial dual-rail encoded dimon qubit, the only difference in architecture from a conventional transmon is in the fabrication of an alternative qubit design, while the physical footprint and control wiring remain unchanged, making the infrastructure significantly more scalable \cite{wills2025error}. \\

Additionally, unlike approaches based on engineering biased Pauli noise, engineering erasure qubits enables immediate advantages for current and near-term devices in the NISQ (Noisy Intermediate Scale Quantum) era. In particular, recent experiments have already demonstrated erasure-detected operations and postselected logical improvements in superconducting platforms \cite{chou2024superconducting, mehta2025bias, wills2025error}. More broadly, a substantial body of work has shown that error detection combined with postselection can act as a powerful form of quantum error mitigation for near-term algorithms \cite{bonet-monroig2018low, temme2017error, koczor2021exponential}. Hence, we expect these approaches to lead to advances in near-term superconducting hardware quantum computers, in addition to enabling accelerated hardware roadmaps for fault-tolerant quantum computing.
\section{Erasure error-correction cycle}\label{section:erasure-errors}

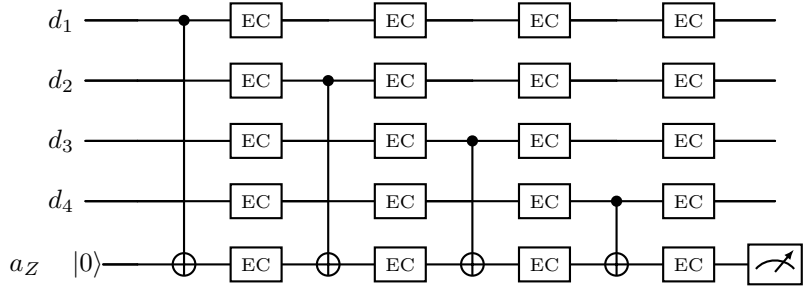
\begin{figure}[h!]
    \centering
    \begin{quantikz}[row sep=0.35cm, column sep=0.45cm]
        \lstick{$d_1$} & \qw & \ctrl{4} & \ec & \qw & \ec & \qw & \ec & \qw & \ec & \qw \\
        \lstick{$d_2$} & \qw & \qw & \ec & \ctrl{3} & \ec & \qw & \ec & \qw & \ec & \qw \\
        \lstick{$d_3$} & \qw & \qw & \ec & \qw & \ec & \ctrl{2} & \ec & \qw & \ec & \qw \\
        \lstick{$d_4$} & \qw & \qw & \ec & \qw & \ec & \qw & \ec & \ctrl{1} & \ec & \qw \\
        \leftlabel{$a_Z$}{\ket{0}} & \qw & \targ{} & \ec & \targ{} & \ec & \targ{} & \ec & \targ{} & \ec & \meter{}
    \end{quantikz}
    \caption{Syndrome extraction circuit for a $d=2$ surface code with integrated erasure checks.  
    The data qubits $d_i$ and ancilla $a_Z$ are dual-rail encoded qubits.  
    Each erasure check flags leakage out of the computational subspace, after which the affected qubit is reset before continuing the circuit.}
    \label{fig:erasure-circuit}
\end{figure}

To see how an erasure error-correction protocol operates in practice, consider the surface-code syndrome extraction circuit in figure~\ref{fig:erasure-circuit}.  
The four data qubits $d_i$ form a surface code, and the ancilla $a_Z$ measures a $Z$-type stabiliser that flags $Z$ errors on the data qubits.  
Both data and ancilla are dual-rail encoded qubits, so erasure checks can be applied throughout the circuit.  \\

Erasure checks are performed after each gate. If a check flags, the corresponding qubit is reset to the computational subspace $\{\ket{01},\ket{10}\}$, a process known as \textit{erasure conversion}. For analysis and simulation, this reset is often treated as producing a maximally mixed state in that subspace, so the resulting error channel is equivalent to a depolarising (Pauli) channel. The stabiliser measurement on $a_Z$ then identifies the error type, while the erasure flag identifies the location. Together, these two pieces of information make the error fully correctable. \\ 

More broadly, incorporating erasure-derived location information into the decoder for any outer code makes the errors more readily correctable.  
By conditioning on which qubit (and when) the event occurred, the decoder rules out many otherwise-degenerate error patterns, improving the accuracy of its predictions (see section \ref{subsec:decoding}).  \\

In realistic devices, erasure checks may be imperfect. False negatives can miss erasures, while false positives can over-flag clean qubits, introducing unnecessary additional errors. The optimal frequency and placement of checks depend on the noise model, as explored in~\cite{chang2024surface,gu2024optimizing} and discussed further in section~\ref{section:open-questions}.  

\subsection{Comparison with other noise channels}\label{subsec:surface-example}

\begin{figure}[h!]
    \centering
    \includegraphics[width=0.8\linewidth]{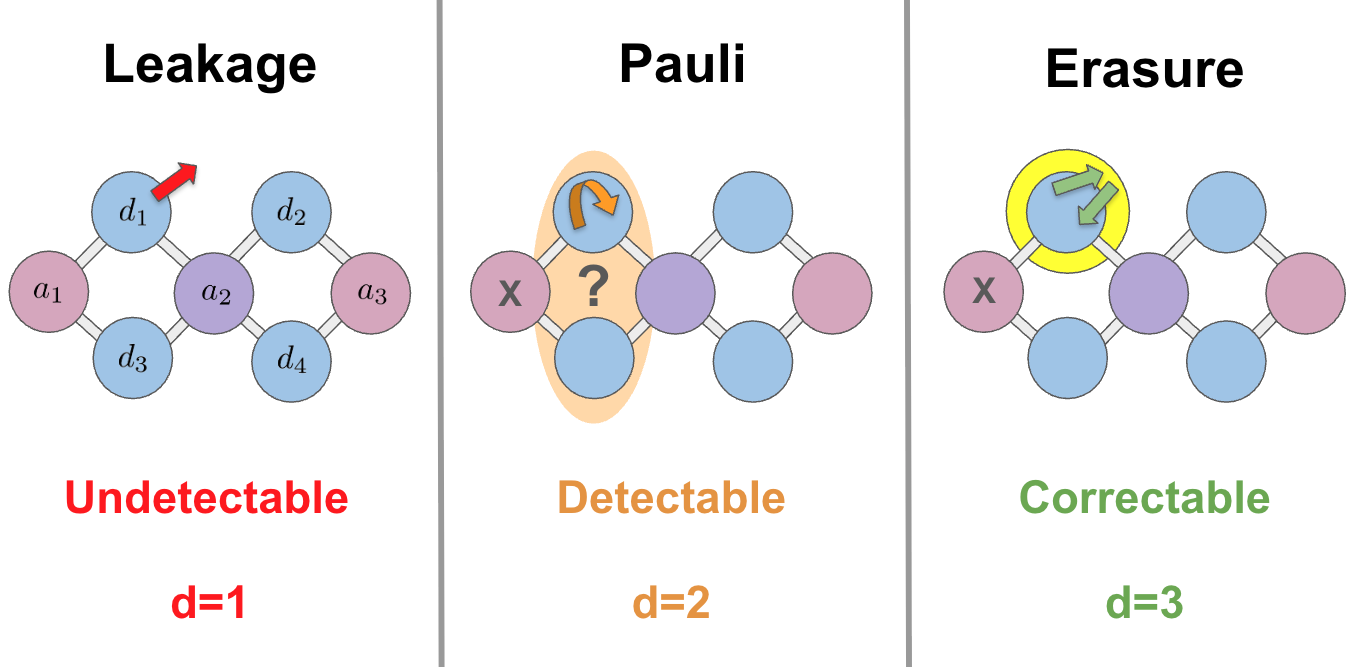}
    \caption{Comparison of leakage, Pauli, and erasure errors in a minimal surface-code layout, with data qubits $d_i$, ancilla qubits $a_i$, and effective code distance d. Leakage errors escape the computational subspace without any flag (undetectable, distance $d=1$);  
    Pauli errors stay within the subspace and flip stabilisers (detectable, $d=2$);  
    erasure errors initially escape the computational subspace, then are flagged by erasure checks and can be reset and corrected (correctable, $d=3$).}
    \label{fig:error-types}
\end{figure}

The $n=4$ surface code example also clearly illustrates the hierarchy of erasure, Pauli and leakage errors, which each have a different effective distance and resulting protection against errors, depicted in figure~\ref{fig:error-types}. \\

In addition to the four data qubits, $d_1$–$d_4$, the figure shows three ancilla qubits; two to check for X-errors on the computational states of these data qubits (bit-flips) and one to check for Z-errors (phase-flips). Specifically, in this example ancilla $a_1$ checks for $X$-errors on $d_1$ or $d_3$; $a_2$ checks for $Z$-errors on $d_1$, $d_2$, $d_3$, or $d_4$; and $a_3$ checks for $X$-errors on $d_2$ and $d_4$.  
The ancillas begin in the state $\ket{000}$, and their measurements yield a three-bit syndrome pattern that reveals which stabilisers were violated.  \\

For a single Pauli error, the ancillas that change state indicate whether the error was of $Z$-type (via $a_1$ or $a_3$, syndromes 100 or 001), $X$-type (via $a_2$, syndrome 010), or $Y$-type (via both $a_2$ and either $a_1$ or $a_3$, syndromes 110 or 011). Hence, for arbitrary Pauli noise, a single error can always be detected; however, there is ambiguity about which data qubit produced that syndrome, since multiple error locations share the same stabiliser check outcomes.\\

When the underlying noise is instead erasure noise, each erasure check directly identifies the data qubit $d_i$ that left the computational subspace.  
The syndrome then provides the error type while the erasure flag gives the location, resolving this ambiguity and enabling full correction of the error. \\ 

In summary: leakage errors leave the computational subspace without any detectable flag, making them undetectable ($d=1$).  
Pauli errors stay within the subspace and flip stabilisers, making them detectable but not localisable ($d=2$).  
Erasure errors are both detectable and localisable. When an erasure check flags, the qubit’s location is known, and after erasure conversion the resulting Pauli-type error can be diagnosed and corrected by the stabilisers.  
The effective distance for erasure errors is therefore $d=3$, such that they are correctable.  

\subsection{Decoding erasures}\label{subsec:decoding}

A key component of correcting errors is to take the classical information from a syndrome measurement, and decode to find out the most likely errors. This is done using a classical decoding algorithm. There are various decoding algorithms being developed and optimized, with some of the most efficient and popular ones for the surface code being the Minimum Weight Perfect Matching (MWPM) and Union Find (UF) decoders. \\

When considering the task of decoding for a surface code with erasure errors, a natural direction is to modify the standard surface code decoding algorithms, such that they include the additional information from erasure checks about the locations at which the errors happened. For example, this has been done with a weighted UF decoder in \cite{wu2022erasure}, and with MWPM in \cite{kang2023quantum}, using the efficient stabilizer simulator stim \cite{gidney2021stim}. Note that the UF decoder is optimal for pure erasure errors (models with only erasure noise, and no Pauli noise) \cite{delfosse2020linear, delfosse2021almost}. \\

A related development is work based on updating a decoder in real-time based on heralded leakage events, explored in the recently introduced Local Clustering Decoder which performs leakage-aware decoding \cite{ziad2024local}. While the motivation of the decoder is to reduce the impact of errors from unintended leakage, a similar approach can apply for intentionally designed erasure errors. In the leakage-aware decoder, the location information is incorporated using the Union-Find approach, which decodes based on \textit{clusters} where errors are predicted to take place. The location information is used to pre-condition the locations of those clusters, resulting in increased accuracy of the decoder. The general erasure decoding problem has also been recently analysed in \cite{kuo2024degenerate}. \\

We have seen that erasure errors are therefore advantageous over pure Pauli errors; if both are happening at the same rate, then there is a clear advantage to having the erasure error with location information compared to having the Pauli error. To maximally exploit this, erasure errors must be the dominant type of error, as discussed in section \ref{subsec:hierarchy} on the hierarchy of errors. 

\section{Superconducting erasure qubit implementations} \label{section:implementations}

\begin{figure}
    \centering
    \includegraphics[width=1\linewidth]{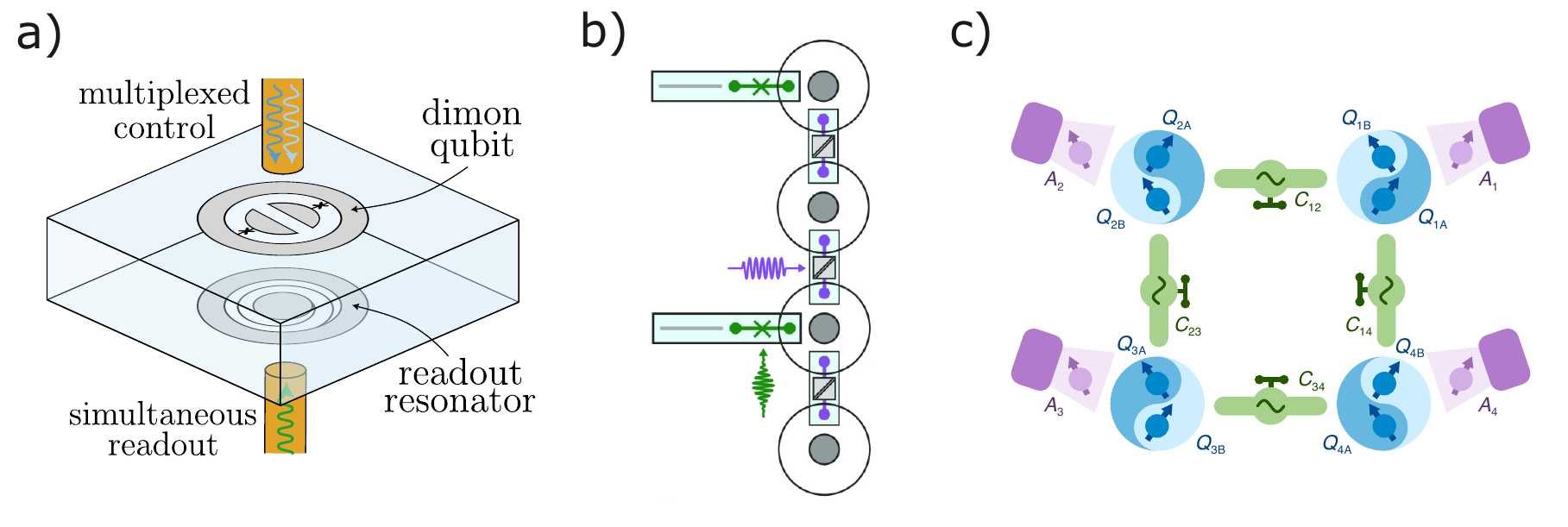}
    \caption{Three example implementations of superconducting erasure qubits: a) coaxial dimon; b) cavity QED; c) coupled transmons. Panels b) and c) are reproduced from \cite{teoh2023dual}, figure~4, and \cite{huang2025logical}, figure~1b, respectively; see the original works for further details.}
    \label{fig:erasure_qubits}
\end{figure}

There are several ways to realise a superconducting erasure qubit, where the dominant noise is erasure errors rather than Pauli errors.  
Three main approaches have been explored to date, with experimental demonstrations, all of which use a dual-rail encoding. These are: coupling two transmons, engineering multimode (``dimon'') qubits, and using cavity QED systems, each depicted in figure \ref{fig:erasure_qubits}.\\

\textbf{Coupled transmons and ancillas.} One way to create a dual-rail qubit that spans the states $\ket{01}$ and $\ket{10}$ with superconducting hardware is to encode it as a two-qubit code using two coupled transmon qubits. Then, the standard amplitude damping error whereby a qubit goes from $\ket{1}$ to $\ket{0}$ becomes an erasure error, taking the two-qubit state to $\ket{00}$ if either physical transmon has an amplitude damping error.  Bit-flips in the encoded erasure qubits, i.e. transitions such as $\ket{01}$ to $\ket{10}$, are suppressed because that would involve a two-photon transition, where one qubit de-excites and the other excites.\\

It is not immediately clear that this encoding helps compared with using the physical transmons directly. With a bare transmon, only the $\ket{1}$ state relaxes while $\ket{0}$ does not. By contrast, in the dual-rail encoding with $\ket{01}$ and $\ket{10}$, each logical state contains one excitation, so either physical transmon can relax. As a result, erasures occur at roughly twice the amplitude-damping rate of a single transmon. Nevertheless, the ability to locate each event can more than compensate for this, improving logical performance under surface-code decoding \cite{kubica2023erasure}.\\

Even after amplitude damping errors have been converted into detectable erasures, there remains a dephasing error. There are standard ways to suppress dephasing, most notably dynamical decoupling \cite{yan2018distinguishing, burnett2019decoherence} .  With amplitude damping converted into erasures, the (significantly smaller) dephasing error and the quality of quantum coherent control become the new limiting factors for logical performance \cite{kubica2023erasure}. This makes robust dual-rail gate design increasingly relevant \cite{singhal2025robust}.\\

The coupled-transmon approach has been demonstrated by Caltech, AWS, and Shenzhen groups \cite{kubica2023erasure, levine2024demonstrating, huang2025logical}, and offers a path to building erasure-biased qubits using standard transmon technology.\\

\textbf{Multimode superconducting qubits.} A multimode qubit augments the transmon with an extra island and junction, creating two transmon-like modes, conventionally known as a \textit{dimon}. Where a transmon qubit consists of two superconducting islands and one Josephson junction, the dimon consists of three superconducting islands, and two Josephson junctions. The dual-rail logical qubit is encoded in the single-excitation subspace, so relaxation in either mode drives the system to $\ket{00}$ and can be heralded as an erasure event. This implementation has a naturally high logical $T_2$ time arising from the fact that the $\ket{01}$ and $\ket{10}$ states are sensitive to similar noise sources. As a result, there is low relative dephasing between $\ket{0}_L$ and $\ket{1}_L$. \\ 

Researchers at OQC construct this device in a coaxial circuit QED architecture, where control and readout is achieved via out-of-plane coaxial control lines \cite{rahamim2017double}. This fixed-frequency approach reduces unwanted noise, crosstalk, and calibration overheads. The dual-rail encoded dimon qubit has been demonstrated to show improved coherence over physical modes whilst retaining the same physical footprint of the coaxmon architecture \cite{wills2025error}. The extensible architecture presents a path to large scale quantum processors without the complexity of additional ancillary hardware and wiring. \\

\textbf{Cavity QED.} Another approach which has seen multiple recent developments from academia and industry is to use superconducting qubits based on cavity QED (cQED) to implement dual-rail qubits. The motivation for using cQED over transmons is greater coherence times and lower intrinsic dephasing rates. Recent results from groups at Yale and Quantum Circuits have demonstrated key aspects required for performing error-correction with erasure qubits, such as mid-circuit erasure detection and the desirable error hierarchy dominated by erasure errors followed by Pauli and leakage errors \cite{chou2024superconducting, koottandavida2024erasure,
de2025mid}. There is also a demonstration of two-qubit gates that preserve the error hierarchy, showing that the noise biased towards erasures can be maintained when applying operations on qubits \cite{mehta2025bias}.  \\

\subsection{Alternative erasure-qubit designs} \label{subsection:alternative-qubits}

The mentioned examples are far from the only ways that qubits can be engineered to implement erasure errors. One of the initial proposals for engineering erasure errors with superconductors was to take advantage of qutrits: use one of the qutrit levels as the state for erasure detection, and the other two as the computational subspace \cite{kubica2023erasure}.\\

Another approach that was mentioned in the same work is the idea of encoding a four-qubit error-detecting code instead of just a two-qubit dual-rail qubit as the inner code, possibly enabling even more powerful gains from a code-concatenation with a outer code. However, in that case the conclusion was that the additional complexity of engineering such a qubit would outweigh the benefits. There is also a recent proposal for implementing a superconducting erasure qubit using a Floquet fluxonium molecule \cite{thibodeau2024floquet}.\\

More generally, there is lots of space for engineering improvements to the current first iterations of erasure qubits on superconductors, through both continuous improvements to the currently developed ones, and exploring radically different structures like those mentioned above. The vast space of possible designs can be curated and constrained by further research on which noise profiles combined with which outer codes provide the optimal gains from erasure errors.\\

\subsection{Applications to error-detection, mitigation and early fault-tolerance} \label{section:NISQ}

While the full gains of erasure qubits are in the context of increasing threshold for fault-tolerant quantum computers, there are also clear applications of heralded erasures to quantum error detection as a near-term error-mitigation technique. Recent superconducting demonstrations of erasure-detected operations already highlight these benefits \cite{chou2024superconducting, mehta2025bias, wills2025error}. Performing erasure checks at the end of a circuit and post-selecting on runs without detected erasures can significantly improve effective coherence times, even before encoding the erasure qubits in an outer code. This strategy is consistent with broader NISQ results showing that error detection and postselection can strongly suppress noise in near-term algorithms \cite{bonet-monroig2018low, temme2017error, koczor2021exponential}, and could improve the implementation of common algorithmic primitives such as the quantum Fourier transform on current devices. \\

The high overheads of post-selection, caused by having to run many shots to obtain useful data, lead to known challenges when scaling error-mitigation techniques (see, e.g., \cite{takagi2023universal, filippov2024scalability}). However, even in the context of full quantum error correction, the quantum error-detection primitive remains valuable; for instance, it can mitigate errors in ancilla qubits, which are frequently measured and often limit logical performance. Regardless of how far error detection is ultimately used to support NISQ applications or to optimise aspects of fault-tolerant architectures, it is intrinsically a required stepping-stone towards erasure-based error correction. Hence, any near-term benefits are effectively a bonus on the path toward the long-term goal of fault tolerance.\\

\section{Open questions and avenues for progress} \label{section:open-questions}

We have discussed several open questions which could lead to interesting further explorations in the near term. These include the possibility of modifying hardware designs to further optimise the advantage to be gained from engineering erasure noise, discussed in section \ref{subsection:alternative-qubits}, and the unknowns regarding which codes stand to gain most significantly from enabling erasure-based error correction. \\ \\

\textbf{Optimisations}\\

There are also various other directions for optimisations that have begun to be explored in the literature. For example, one of these explored in \cite{gu2024optimizing} is the relevance of the protocol used for the reset of an erased qubit back to the computational subspace. This can for instance can be a \textit{unitary pulse} (which risks taking a good qubit outside the computational subspace in the case of a false-positive erasure check) or a \textit{one-way pulse} which sends any state to the computational subspace. Another degree of freedom explored in \cite{gu2024optimizing} is optimising the frequency of erasure checks, trading clock time (which increases risk of idle errors) for additional information about erasures.  A further area for investigation is having more detailed noise models to understand the effect of imperfections in the erasure checks, and the trade-offs involved, analysed in \cite{chang2024surface}. \\

There may also be optimisations possible regarding efficient decoding of the error patterns from erasure qubits. This will depend on the nature of the outer code, since the approach to decoding can differ vastly for different codes. Further developments of results based on detecting and mitigating leakage errors (e.g. \cite{ziad2024local}) may also have application to erasure error decoding, given that they also deal with errors produced by leaving the computational subspace that are detected and then reset.\\

\textbf{Software tools for erasures}\\

Erasure-based quantum error-correcting codes differ both structurally and practically from conventional codes that do not incorporate erasure information. These distinctions pose significant challenges for the direct application of existing design~\cite{crumble, loom} simulation~\cite{gidney2021stim, riverlane_qecx}, decoding~\cite{decoding} and resource estimation~\cite{azure_resource, bartiq, qualtran} tools, which are typically designed for depolarizing or Pauli noise models without erasure flags. As such, existing tooling need extensions and adaptations for erasure-based codes. Adaptation of existing tools and development of new ones focused on code concatenation and erasure codes is essential for accelerated development of the field.\\

In particular, tools that support flexible code concatenation schemes, hybrid error models, and erasure-aware decoding strategies are critical to unlocking the full potential of erasure codes. Leveraging information about the location of errors requires custom simulation pipelines, decoders, and performance metrics. The development of such specialized tooling will not only facilitate rigorous benchmarking of erasure codes but also accelerate their integration into practical fault-tolerant quantum computing architectures.\\

\section{Conclusion}

The bottleneck for achieving useful quantum computing with superconducting qubits is the short T$_1$ times, caused by amplitude damping errors \cite{kubica2023erasure}. Importantly, the physical error rate must be below threshold for quantum error correction in order to enable large-scale fault-tolerant quantum computing. Recent developments with dual-rail superconducting erasure qubits give a promising solution: by engineering the noise of superconducting qubits to be detectable erasures, the dominant source of error in superconducting qubits can be efficiently corrected. The threshold error can be significantly increased, up to realistically achievable rates with current and near-term qubits. In addition, logical error with detectable erasures is reduced, meaning a lower ratio of physical qubits to logical qubits is required for achieving a given code distance. \\

Recent simulations and proof-of-principle demonstrations show that these theoretical advantages carry over to realistic hardware. The erasure-qubit approach to QEC is now being explored across superconducting, neutral-atom and trapped-ion systems, each leveraging the dual-rail encodings that originate in photonic devices.\\

The combination of theoretical advances and the maturity of today's superconducting qubits has fuelled the rapid transition from theory to simulation to demonstration of these techniques. The close relationships of companies and academic research groups have also enabled new ideas from academic labs to be adopted and developed in industry settings with fast turnaround times. However, even with this momentum, the implementation of erasure qubits remains in its infancy, with many open questions left to explore in order to characterise its full potential and take advantage of the various new degrees of freedom introduced by this approach.\\

These early results are already shaping how the field thinks about scalable quantum computing architectures. The quantum industry recognises that these devices require strategic co-design of hardware with quantum error correction, with features such as a large logical-to-physical qubit ratio; high thresholds to physical noise; and architectures whose complexity does not grow prohibitively as systems scale. Engineering dual-rail superconducting erasure qubits provides an avenue to achieving these goals for early fault-tolerant devices. It is also a demonstrative example of the innovation possible on future qubit designs, building on the principle of concatenating error-correcting properties of hardware with outer codes to directly address the dominant physical noise sources.\\

Erasure qubits, and more generally hardware-efficient QEC, offers a credible route to lower-overhead, higher-performance logical qubits. They therefore represent one of the most promising directions for advancing toward scalable, fault-tolerant quantum computing.

\section*{Acknowledgements}
The authors thank Ailsa Keyser, William Howard, Ofek Asban and Ben Rogers for discussions and review of this manuscript. We thank the OQC team for their support.
\printbibliography

\end{document}